\documentclass[fleqn,usenatbib]{mnras}

\usepackage[T1]{fontenc}
\usepackage[utf8]{inputenc}
\usepackage{newtxtext,newtxmath}
\usepackage{graphicx}
\usepackage{amsmath}
\usepackage{booktabs}
\usepackage{placeins}
\usepackage{dblfloatfix}
\usepackage{lineno}
\graphicspath{{./population_new/figure/}}
\hypersetup{colorlinks=true,citecolor=blue,linkcolor=blue,urlcolor=blue}

\title{Binary Progenitors of Electron-capture Supernovae}

\author[Zhong \& Shao]{
Liji Zhong$^{1}$,
Yong Shao$^{1,2}$\thanks{E-mail: shaoyong@nju.edu.cn}\\
$^{1}$Department of Astronomy, Nanjing University, Nanjing 210023, People's Republic of China\\
$^{2}$Key Laboratory of Modern Astronomy and Astrophysics, Nanjing University, Ministry of Education, Nanjing 210023, People's Republic of China
}

\date{Accepted XXX. Received YYY; in original form ZZZ}
\pubyear{2026}

\DeclareRobustCommand{\revadd}[1]{\ifmmode\boldsymbol{#1}\else{\bfseries\boldmath #1}\fi}
\begin{document}
\raggedbottom
\label{firstpage}
\pagerange{\pageref{firstpage}--\pageref{lastpage}}
\maketitle

\begin{abstract}
Electron-capture supernovae (ECSNe) arise from a narrow progenitor parameter space between oxygen-neon white dwarf formation and iron-core collapse supernovae, governed by the competition between core growth and envelope removal. Binary interactions modify this balance through mass transfer and mergers. 
Using the detailed binary population synthesis framework POSYDON, we model ECSN progenitors across metallicities from $Z/Z_\odot=2$ to $10^{-4}$, focusing on the initial parameter space, pre-supernova stripping, envelope-based classification, surviving companions, post-supernova orbits, and cosmic event rates. The lower edge of the binary ECSN initial primary mass lies near $7\,M_\odot$, owing to efficient convective overshooting in the POSYDON grids, whereas the upper edge increases with metallicity because stronger line-driven winds require larger initial masses to build carbon-oxygen cores within the ECSN range. Stable mass transfer and mergers occupy distinct regions in initial mass ratio and orbital period space. ECSNe occurring in bound binaries are divided into hydrogen-rich Type II ($\sim30\%$), IIb ($\sim10\%$), and Ib ($\sim60\%$) outcomes according to residual hydrogen and helium envelopes. When all progenitor channels are considered, ECSNe occurring in binaries that remain bound at the time of explosion account for approximately one-third of the total ECSN population, whereas the remaining two-thirds arise from progenitors without a bound companion at explosion. Among these bound systems, approximately $90\%$ of the companions are main-sequence stars at explosion. These main-sequence companions exhibit a bimodal spin distribution, including a rapidly rotating component associated with prior mass transfer. Evolved-star and compact-object companions each account for only a few percent.  Combining the formation efficiencies with cosmic star formation and metallicity evolution models yields an intrinsic ECSN rate peaking at $z\sim2-3$ and declining at higher redshifts. Over cosmic history, ECSNe contribute approximately $1\%-4\%$ of all core-collapse supernovae. 
\end{abstract}


\section{Introduction}

Electron-capture supernovae (ECSNe) are generally understood to arise from a narrow final-fate boundary between the formation of oxygen-neon white dwarf (ONe WDs) and iron-core collapse supernovae (Fe CCSNe). In single-star evolution at solar metallicity, ECSN progenitors are associated with zero-age main sequence (ZAMS) masses of roughly $8-10\,M_\odot$ \citep{nomoto_evolution_1984,nomoto_evolution_1987}. These stars enter the super-asymptotic giant branch (SAGB) phase and develop highly degenerate ONe cores. Their final fate is determined by the competition between core growth via shell burning during the thermally pulsing SAGB phase and the removal of the hydrogen envelope through mass loss \citep{ritossa_evolution_1999,siess_evolution_2007,poelarends_supernova_2008,doherty_super-agb_2017}. If the core reaches the electron-capture threshold before the envelope is lost, electron captures on $^{24}\mathrm{Mg}$ and $^{20}\mathrm{Ne}$ reduce the electron degeneracy pressure and trigger runaway contraction, leading to collapse and neutron star formation \citep{miyaji_supernova_1980,nomoto_evolution_1984,nomoto_evolution_1987}. Conversely, if envelope stripping occurs earlier, core growth is halted and the star is more likely to leave a massive ONe WD.

Even after the core reaches the electron-capture threshold, collapse to a neutron star is not guaranteed. The outcome depends on the density at which oxygen ignites, as well as on the ignition conditions and the flame physics. At lower central density, thermonuclear energy release can partially reverse the collapse rather than proceeding directly to neutron star formation \citep{leung_electron_capture_2020,holas_drawing_2026}. Multidimensional hydrodynamical simulations by \citet{jones_electron-capture_2016,jones_remnants_2019} have shown that such cases may leave a bound ONeFe WD remnant instead of a canonical neutron star. The ECSN channel thus occupies only a narrow region of progenitor parameter space, reflecting the strong sensitivity of the final outcome to late-stage core structure.

The initial mass interval for ECSNe is sensitive to assumptions about internal mixing. Models with weak or no convective overshooting produce smaller cores at fixed initial mass and tend to place the lower edge of the ECSN window near $9\,M_\odot$. More recent detailed calculations with efficient convective overshooting place this lower edge closer to $7\,M_\odot$ \citep{eldridge_progenitors_2004,siess_evolution_2007,jones_advanced_2013,doherty_super-_2015,doherty_super-agb_2017}. For population synthesis, even a modest shift in the ECSN mass window changes the predicted event rate.

ECSNe are expected to differ observationally from Fe CCSNe. Their steep pre-collapse density gradients lead to relatively weak explosions, low ejecta masses, and small $^{56}$Ni yields, typically $\sim 10^{-3}-10^{-2}\,M_\odot$ \citep{timmes_neutron_1996,kitaura_explosions_2006,wanajo_electron-capture_2013,wanajo_electron-capture_2013-1,wanajo_nucleosynthesis_2018}. The natal kicks of the resulting neutron stars are also commonly assumed to be smaller than those of Fe CCSNe, making ECSNe relevant to the formation of bound neutron star binaries and low-eccentricity high mass X-ray binaries \citep{pfahl_new_2002,podsiadlowski_effects_2004,tauris_formation_2017,shao_formation_2015,Shao2018,Guo2024,Wang2026,Li2026}. ECSNe have also been considered as a possible origin of the low-mass end of the neutron star birth mass distribution. \citet{schwab_further_2010} linked a low-mass component to electron-capture explosions, while more recent reconstructions suggest that the birth mass function remains sensitive to both explosion physics and binary evolution \citep{you_determination_2025}. The proposed ECSN origin of SN 2018zd connects these expectations to an observed event with evidence for a SAGB-like progenitor, dense circumstellar material, low explosion energy, and characteristic nucleosynthetic signatures \citep{hiramatsu_electron-capture_2021}.

A single-star description is insufficient for this problem, given that a large fraction of massive stars are born in binaries and undergo binary interaction before core collapse \citep{sana_binary_2012,moe_mind_2017}. Roche lobe overflow (RLOF), mass accretion, common envelope (CE) evolution, and mergers can all modify the envelope stripping history and hence the growth of the core. For potential ECSN progenitors, early RLOF can strip the envelope before the core reaches the ECSN threshold, redirecting the star toward ONe WD formation. By contrast, later interaction can modify the remaining envelope structure close to collapse, while mergers can increase the effective core mass and bring systems that would otherwise remain below the single-star threshold into the ECSN regime. Binary evolution should affect not only the ECSN rate but also the pre-supernova envelope mass, spectral appearance, surviving companion, and post-supernova orbit \citep[e.g.,][]{podsiadlowski_effects_2004,tauris_ultra-stripped_2015,poelarends_electron_2017}. For stripped-envelope ECSNe, \citet{moriya_rapidly_2016} showed that the low ejecta mass can lead to rapidly evolving faint transients, illustrating how binary stripping may alter the observable appearance.

Systematic ECSN studies that combine detailed stellar structure with broad binary parameter coverage remain scarce. Rapid population synthesis approaches can efficiently explore large samples, but their analytic prescriptions do not always capture the response of the donor, core, and envelope during RLOF \citep{hurley_evolution_2002,Belczynski2008,team_compas_rapid_2022,He2026}. Detailed stellar evolution studies provide a more physical treatment of mixing and stripping, but are usually restricted to smaller grids and often do not simultaneously cover a wide metallicity range together with the main binary initial conditions (primary mass, mass ratio, and orbital period). This limitation matters for ECSNe, where the allowed core structure range is narrow enough that modest changes in mixing, mass loss, or binary interaction history can move a system between ONe WD formation, ECSN and Fe CCSN.

In this work, we use the detailed binary population synthesis framework POSYDON \citep{fragos_posydon_2023,andrews_posydon_2025} to study ECSN progenitors across metallicity. Built on grids of detailed MESA binary calculations \citep{paxton_modules_2015}, POSYDON provides a more physically self-consistent description of the structural response to binary interaction than rapid population synthesis methods. We explore how the ECSN progenitor parameter space varies with metallicity and with the main binary initial conditions: primary mass, mass ratio, and orbital period. We also relate the residual hydrogen and helium envelopes of the pre-supernova models to their expected spectral classes, and connect the surviving companions and post-supernova orbits to their earlier interaction histories. Finally, we combine the formation efficiencies obtained at different metallicities with cosmic star formation and metallicity evolution models to estimate the intrinsic cosmic ECSN rate.

The paper is organized as follows. Section 2 outlines the adopted modelling framework, including the wind and binary interaction prescriptions, the ECSN selection criterion, and the initial population. Section 3 presents the formation channels across metallicity, pre-supernova stripping and classification, companion properties, post-supernova orbits, and cosmic event rates. Section 4 summarizes the main conclusions.

\section{Methods and Models}

To model ECSN progenitors across metallicity and binary initial conditions, we use the POSYDON v2 framework \citep{fragos_posydon_2023,andrews_posydon_2025}, which combines Monte Carlo binary population synthesis with interpolation over precomputed MESA grids. In contrast to rapid population-synthesis codes such as BSE, StarTrack, COMPAS, and POPKIN which typically use analytic prescriptions for stellar structure and binary interaction \citep{hurley_evolution_2002,Belczynski2008,team_compas_rapid_2022,He2026}, POSYDON allows us to follow core growth and envelope stripping through detailed binary evolution calculations. The following sections summarize the adopted wind prescriptions, internal-mixing assumptions, CE treatment, ECSN selection criterion, and population synthesis setup.

\subsection{Stellar structure and binary interactions}

Envelope removal directly affects whether a star remains in the ECSN progenitor regime, making the adopted wind prescription a central part of the model. We use the mass-loss scheme associated with the underlying POSYDON stellar evolution grids. For stars with initial masses $M\ge8\,M_\odot$, we adopt the temperature-sensitive Dutch wind scheme, with the line-driven rates of \citet{vink_mass-loss_2001} applied in hot phases ($T_{\rm eff}>11{,}000\,{\rm K}$) and the empirical prescription of \citet{de_jager_mass_1988} used in cooler red supergiant phases ($T_{\rm eff}<10{,}000\,{\rm K}$). Between these temperatures, the mass-loss rate is linearly interpolated. For stars with $M<8\,M_\odot$, which can still enter the ECSN regime through efficient core growth or binary interaction, the hot-wind treatment is retained for $T_{\rm eff}>12{,}000\,{\rm K}$, whereas the cooler phases below $T_{\rm eff}<8{,}000\,{\rm K}$ are treated with the \citet{reimers_circumstellar_1975} and \citet{bloecker_stellar_1995} prescriptions, using $\eta_{\rm R}=0.1$ on the first giant branch and $\eta_{\rm B}=0.2$ during the thermally pulsing phase, where $\eta_{\rm R}$ and $\eta_{\rm B}$ are dimensionless scaling factors for the corresponding mass-loss rates. The intermediate temperature regime is again handled by linear interpolation. Rotational enhancement of the mass-loss rate follows \citet{langer_coupled_1998}. Unless otherwise specified by the adopted prescription, the wind rate is scaled with initial metallicity following the POSYDON implementation \citep{fragos_posydon_2023}.

Internal mixing affects the ECSN boundary by setting the core mass reached at fixed initial mass. In the POSYDON v2 grids, convection is treated with the Ledoux criterion and a mixing length prescription calibrated to the Sun, while overshoot mixing follows an exponential decay formalism whose strength varies with stellar mass \citep{paxton_modules_2015,fragos_posydon_2023}. The overshooting parameter is set to $f_{\rm ov}=0.016$ for stars below $4\,M_\odot$ and to $f_{\rm ov}=0.0415$ for stars above $8\,M_\odot$\footnote{For $M<4\,M_\odot$, $f_{\rm ov}$ follows the MIST calibration to the main-sequence turnoff morphology of the open cluster M67 \citep{choi_mesa_2016,fragos_posydon_2023}. For $M>8\,M_\odot$, \citet{fragos_posydon_2023} converted the step-overshoot parameter $\alpha_{\rm ov}=0.335$ to the exponential prescription adopted in POSYDON; this parameter was calibrated by \citet{brott_rotating_2011} using rotating massive-star models to reproduce the observed main-sequence width.}, with a smooth ramp between these values over $4-8\,M_\odot$ \citep{choi_mesa_2016,fragos_posydon_2023}. The larger overshooting adopted in the massive-star regime increases the main-sequence convective core and allows stars of smaller initial mass to form larger carbon-oxygen (CO) cores, thereby shifting the lower edge of the ECSN mass range downward. Detailed SAGB calculations also show that the ECSN window is sensitive to the treatment of internal mixing \citep{siess_evolution_2007,jones_advanced_2013,doherty_super-_2015,doherty_super-agb_2017}.

In the detailed MESA binary grids used by POSYDON, ZAMS binaries are initialized with circular orbits and stellar spins synchronized with the orbit
\citep{fragos_posydon_2023}. The population synthesis does not draw an independent initial stellar spin distribution.

CE evolution is treated with the standard $\alpha_{\rm CE}-\lambda_{\rm CE}$ energy formalism \citep{webbink_double_1984}, adopting $\alpha_{\rm CE}=1$. In our POSYDON setup, $\lambda_{\rm CE}$ is taken from the grid-based stellar structure quantities through the \texttt{lambda\_from\_grid\_final\_values} option rather than fixed to a single constant. The core-envelope boundary is defined using hydrogen and helium mass fraction thresholds of 0.3 and 0.1, respectively. For Hertzsprung-gap donors, this boundary is less well defined than in more evolved giants, making it uncertain whether envelope ejection would leave a viable remnant. We adopt the optimistic treatment of Hertzsprung-gap donors, allowing a subset of those entering CE to survive, rather than the pessimistic treatment in which all such donors merge. After successful envelope ejection, the subsequent evolution is followed with the \texttt{two\_phases\_stableMT} option. The envelope is first removed down to the core boundary defined by the adopted abundance thresholds during CE. Any material remaining between this boundary and the helium or CO core in the MESA model before CE is then removed through instantaneous, stable, fully non-conservative mass transfer. These assumptions mainly affect the survival of relatively low-mass binaries and the relative contributions of the stable RLOF and merger channels.

Not all CE episodes lead to successful envelope ejection and binary survival; in some cases, the inspiral ends in merger. In POSYDON v2, the merger product is assigned approximate total and core properties using the layer-summing approach described by \citet{andrews_posydon_2025}, based on \citet{hurley_evolution_2002}, and is then matched to the nearest model in the single-star grids so that its subsequent evolution can be followed within the population synthesis framework. This treatment allows merger products to contribute explicitly to the ECSN population.

\subsection{ECSN criteria and explosion dynamics}

The terminal core-collapse outcome is determined from the pre-collapse stellar structure. For core-collapse remnants in general, we adopt the delayed engine of \citet{fryer_compact_2012}. ECSNe are then identified separately using the CO core criterion of \citet{tauris_ultra-stripped_2015}, rather than an approximate classification based on helium core mass \citep{hurley_evolution_2002}. A system is classified as an ECSN progenitor when its CO core at carbon depletion lies in the range $1.37\,M_\odot \le M_{\rm CO} \le 1.43\,M_\odot$. Systems below this interval are assumed to avoid collapse and leave ONe WDs, whereas systems above it are assumed to proceed to more advanced burning stages and lead to Fe CCSNe. Detailed studies of ONe core ignition and the neon-burning boundary support an ECSN window in this range \citep{siess_evolution_2007,jones_electron-capture_2016,jones_remnants_2019}. \citet{poelarends_electron_2017}, however, noted that efficient cooling after carbon depletion could suppress neon ignition and allow some cores as massive as $1.52\,M_\odot$ to collapse through electron capture. We do not adopt this higher mass extension of the ECSN window, given that the population outputs do not retain enough thermal information to evaluate the relevant cooling history on a case-by-case basis.

Within this supernova setup, we set the maximum allowed neutron star mass to $2.5\,M_\odot$ and cap the neutrino-carried mass loss during proto-neutron star formation at $0.5\,M_\odot$. The neutron star mass limit is a model boundary within the range permitted by uncertainties in the equation of state \citep{kalogera_maximum_1996,alsing_maximum_2018}. This limit is applied to the gravitational mass of the remnant, obtained from its baryonic mass using the relation of \citet{lattimer_analysis_1989} as modified by \citet{zevin_exploring_2020}. The modification assumes a maximum iron core mass of $5\,M_\odot$, with the neutrino mass loss limit set to 10\% of that value. Natal kicks are included for all core-collapse events. For neutron stars formed in Fe CCSNe, we adopt a Maxwellian kick distribution with one-dimensional dispersion $\sigma=265\,{\rm km\,s^{-1}}$ \citep{hobbs_statistical_2005}. For black holes, kick velocities are drawn from the same distribution but scaled inversely with the remnant mass, following the prescription adopted in POSYDON \citep{fragos_posydon_2023}. For ECSNe, we adopt a smaller kick distribution with $\sigma=20\,{\rm km\,s^{-1}}$, motivated by the fast explosion timescale and low ejecta mass expected for ONe core collapse \citep{kitaura_explosions_2006,dessart_multidimensional_2006,gessner_hydrodynamical_2018,tauris_j04531559_2019}. These assumptions mainly affect which binaries remain bound after explosion and shape the companion survival fraction and the post-supernova orbital distributions discussed in Section 3.

\subsection{Population synthesis and metallicity grids}

We perform Monte Carlo binary population synthesis over eight metallicities: $Z/Z_\odot=2.0$, $1.0$, $0.45$, $0.2$, $0.1$, $0.01$, $0.001$, and $0.0001$. At each metallicity, we independently sample $10^5$ stellar systems in each of three initial primary mass intervals ($4-6\,M_\odot$, $6-12\,M_\odot$, and $12-130\,M_\odot$). Initial primary masses ($M_{1,\rm i}$) are drawn from the \citet{kroupa_variation_2001} initial mass function, while initial secondary masses are assigned using the flat initial mass ratio option in POSYDON, with $q_{\rm i}=M_{2,\rm i}/M_{1,\rm i}$. The binary fraction follows the prescription of \citet{moe_mind_2017}; systems not assigned as binaries by this prescription are evolved as initially single stars and retained in the population statistics. For binaries, the initial orbital period ($P_{\rm orb,i}$) distribution follows the extended \citet{sana_binary_2012} prescription over $0.75\le P_{\rm orb,i}/{\rm d}\le6000$.

ECSN progenitors occupy a narrow and intrinsically rare region of initial-mass space. Sampling the three intervals separately with equal numbers of systems therefore provides sufficient statistics to resolve the ECSN-forming region and its formation channels, while retaining coverage of the full primary-mass range. Because this sampling scheme does not follow the relative abundances implied by the initial mass function, each event in interval $k$ at metallicity $Z$ is assigned a statistical weight $w_k(Z)=1/M_{{\rm formed},k}(Z)$, where $M_{{\rm formed},k}(Z)$ denotes the total initial stellar mass represented by the normalized sample. All efficiencies, fractions, and cosmic rates that combine mass intervals are evaluated using these weights.

The lower sampling boundary at $4\,M_\odot$ lies just above the lower edge, approximately $3.9\,M_\odot$, of the detailed hydrogen-rich POSYDON grids \citep{andrews_posydon_2025}. It is therefore set by the available grid coverage rather than by a universal lower-mass threshold for ECSN formation. 
The $6-12\,M_\odot$ interval is chosen to concentrate the sampling on the mass range most relevant to ECSNe produced by the initially more massive star, while maintaining high resolution in the parameter space of interest. At larger initial primary masses, the initially less massive secondary can instead produce an ECSN after mass transfer, rejuvenation, or CE evolution.  

We classify ECSN events according to the system state immediately before explosion into four groups: ECSNe in bound binaries, initially single stars, single stars produced by binary mergers, and single stars from binaries disrupted by an earlier supernova. The bound-binary group includes both primary and secondary ECSNe. At each metallicity, the fractions of these four groups are normalized by the total weighted count of ECSN events.

\section{Results and Discussion}

The ECSN population across metallicity is described by following the problem from formation to observables. We begin with the initial parameter space that gives rise to ECSN progenitors, and then examine how binary stripping shapes their pre-supernova structure, mass loss, and spectral appearance. The surviving companions and post-supernova orbital properties are considered next, as they retain a record of the earlier interaction history. We also quantify the relative contributions of primary and secondary progenitors and identify the channels that produce ECSNe without a bound companion. Finally, these results are placed in a broader context through convolution with cosmic star formation and metallicity evolution models.

\subsection{Formation channels and progenitor parameter space}

\begin{figure}
    \centering
    \includegraphics[width=0.95\columnwidth]{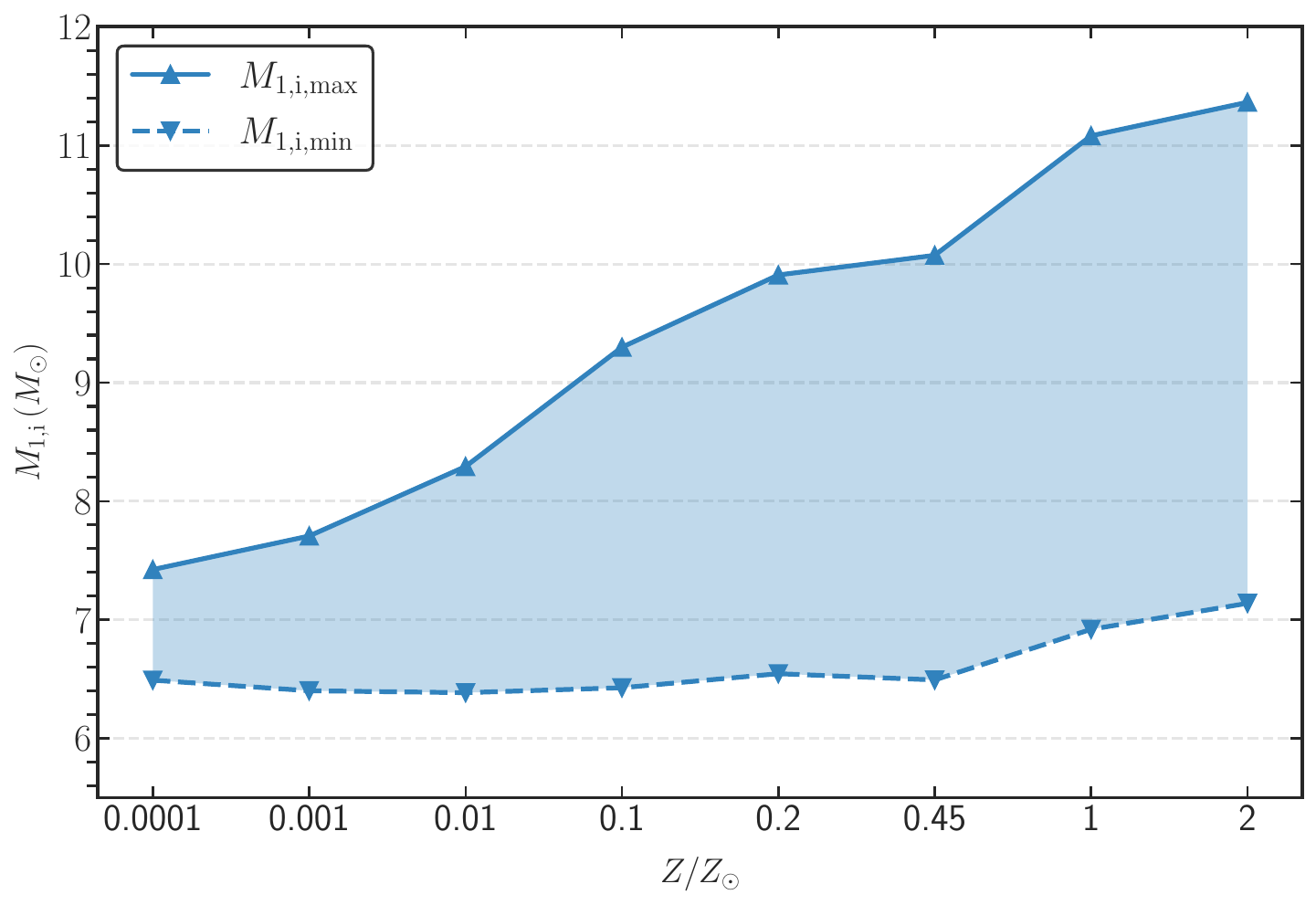}
    \caption{Initial mass range for primary ECSN progenitors in binary systems as a function of metallicity. The lower and upper boundaries are defined by the 1st and 99th percentiles, respectively, of the ECSN-forming primary mass distribution at each metallicity, minimizing the impact of rare outliers.}
    \label{fig:mass_window}
\end{figure}

\begin{figure*}
    \centering
    \includegraphics[width=0.95\textwidth]{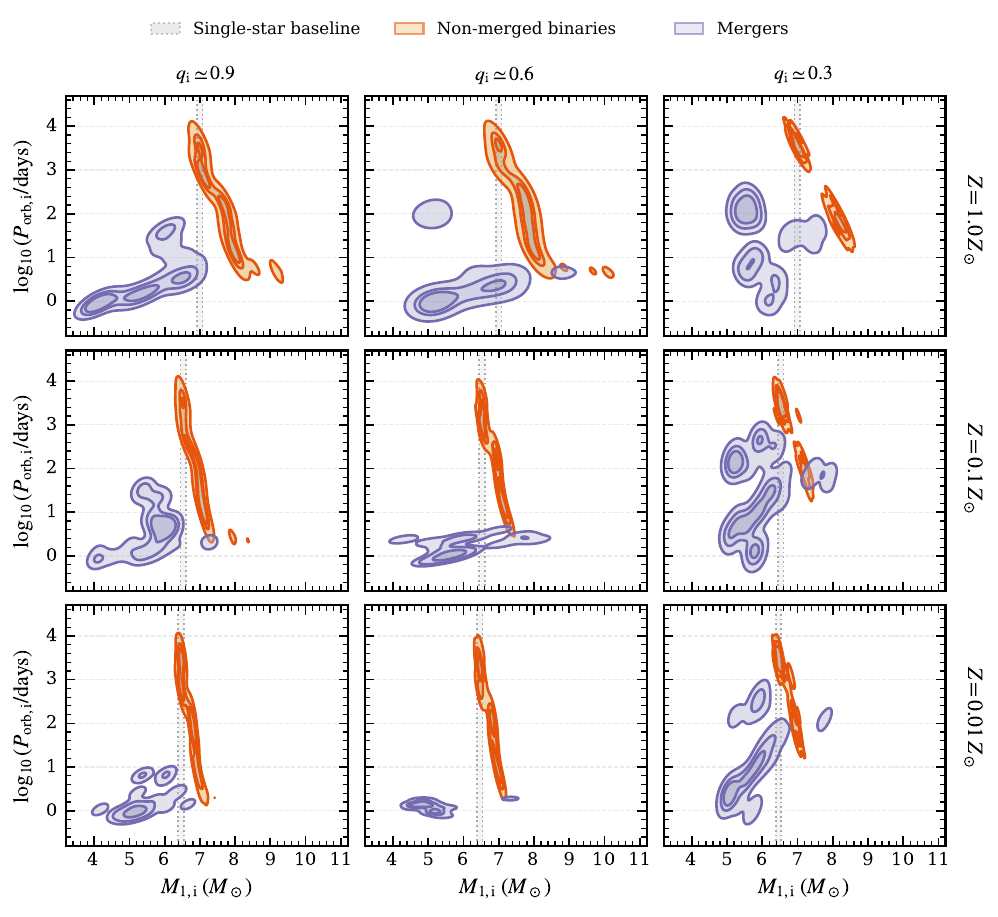}
    \caption{Initial primary mass ($M_{1,\rm i}$) versus initial orbital period ($P_{\rm orb,i}$) parameter space for primary ECSN progenitors. Rows correspond to metallicities $Z=1.0$, $0.1$, and $0.01\,Z_\odot$ (top to bottom), and columns correspond to $q_{\rm i} \simeq0.9$, $0.6$, and $0.3$ (left to right). The light gray shaded band indicates the single-star reference interval. Orange and purple contours represent kernel density estimates for non-merged binaries (i.e., those that avoid merger until primary collapse) and merger systems, respectively.}
    \label{fig:initial_parameter_space}
\end{figure*}

Starting from the birth parameters of primary ECSN progenitors, Fig.~\ref{fig:mass_window} shows the binary ECSN initial primary mass window as a function of metallicity before the full initial parameter space is examined in Fig.~\ref{fig:initial_parameter_space}. Relative to the familiar single-star range of $8-10\,M_\odot$ often quoted for ECSN progenitors, the window in our binary models shifts to lower masses and responds asymmetrically to $Z$.

Across the metallicity grid, the lower and upper boundaries of the binary ECSN window behave differently. The lower edge remains close to $6.5-7.0\,M_\odot$ over most of the metallicity range, whereas the upper edge rises from about $7.5\,M_\odot$ at the lowest metallicities to about $11\,M_\odot$ in metal-rich environments. The overall downward shift relative to the classical single-star expectation is primarily a consequence of internal mixing. In the POSYDON grids, efficient convective overshooting that increases with mass allows stars around $7\,M_\odot$ to develop CO cores large enough to satisfy the \citet{tauris_ultra-stripped_2015} ECSN criterion.

The asymmetric response of the two boundaries to metallicity is driven mainly by stellar winds. Metallicity-dependent line-driven winds are particularly sensitive to luminosity \citep[e.g.,][]{vink_mass-loss_2001}, so the initially more massive and luminous progenitors near the upper edge of the ECSN window tend to undergo stronger mass loss. At higher $Z$, stronger winds make it more difficult for these progenitors to retain the conditions required for ECSN formation, shifting the upper boundary toward larger initial masses.

By contrast, the lower boundary changes much more weakly over most of the metallicity range. Systems near the lower edge have undergone little envelope stripping and are more closely associated with wider binaries. From $Z/Z_\odot=0.0001$ to $0.45$, the minimum $M_{1,\rm i}$ remains close to $6.5\,M_\odot$. These progenitors are less luminous and thus drive weaker line-driven winds, so changes in metallicity have only a limited effect on their cumulative envelope loss. The lower boundary begins to rise only at high metallicity, reaching about $7.0-7.2\,M_\odot$ at $Z/Z_\odot=1-2$, indicating that wind-driven stripping begins to affect even the low-mass edge of the ECSN-forming population.

The resulting mass window is set by two effects, with convective overshooting lowering the mass scale relative to single-star expectations and metallicity-sensitive winds mainly moving the upper boundary, which is much more responsive to $Z$ than the lower boundary. This distribution is expanded across initial orbital period and mass ratio in Fig.~\ref{fig:initial_parameter_space}.

Fig.~\ref{fig:initial_parameter_space} maps the mass interval of Fig.~\ref{fig:mass_window} onto the $M_{1,\rm i}$--$P_{\rm orb,i}$ plane, where ECSN progenitors occupy a confined region that cannot be described by primary mass alone. The light gray band marks the corresponding single-star reference interval, so systems lying close to it can be regarded as evolving nearly as single stars. In this plane, the initial orbital period mainly determines when interaction begins: overly tight orbits initiate mass transfer early enough that envelope removal suppresses subsequent core growth and shifts ECSN production toward larger initial primary masses, while sufficiently wide systems remain close to the single-star case. ECSN formation is concentrated at intermediate periods, for which interaction occurs late enough to allow the core to grow into the electron-capture mass range, but still early enough to prevent further evolution toward Fe core collapse \citep[e.g.,][]{podsiadlowski_effects_2004,poelarends_electron_2017}.

Fig.~\ref{fig:initial_parameter_space} also shows that the distribution of ECSN progenitors varies across the three initial mass ratio ranges. The three labelled $q_{\rm i}$ values correspond to subsamples selected within $|q_{\rm i}-q_{\rm i,0}|\leq0.05$, with $q_{\rm i,0}=0.9, 0.6,$ and $0.3$. We refer to binaries that avoid merger up to primary ECSN as non-merged binaries, a category that includes CE survivors and stable RLOF systems. Merged systems are defined as binaries that undergo a merger before the primary collapses, with the resulting merger product subsequently satisfying the adopted CO-core criterion under the approximate POSYDON treatment. For $q_{\rm i}\simeq0.9$ and $q_{\rm i}\simeq0.6$, the non-merged binaries occupy a continuous band in the $M_{1,\rm i}-P_{\rm orb,i}$ plane at all three metallicities. At $q_{\rm i}\simeq0.3$, this band narrows, while merged systems appear as separated islands at lower primary masses and intermediate periods. The merger component is most prominent at low $q_{\rm i}$, shifting toward lower masses and shorter periods as $q_{\rm i}$ increases. We note, however, that the merger component should be interpreted only qualitatively in the present work, as POSYDON v2 follows merger products through approximate matching to single-star grids.

The trend with metallicity identified in Fig.~\ref{fig:mass_window} is reflected here as a systematic shift of the ECSN-forming region along the primary mass axis. At lower metallicity, the region of binaries that avoid merger extends to smaller $M_{1,\rm i}$, consistent with weaker wind-driven stripping. Toward the metal-rich end, the ECSN-forming region moves to higher initial masses, while its low-mass side no longer extends as far downward.

\begin{figure*}
    \centering
    \includegraphics[width=0.98\textwidth]{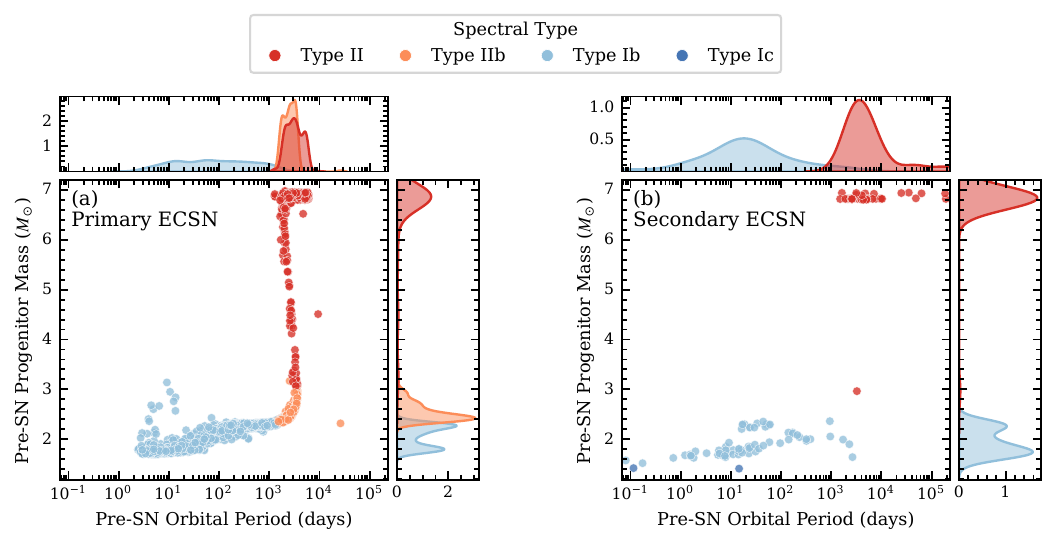}
    \caption{Joint distribution of pre-supernova orbital period and progenitor mass for (a) primary and (b) secondary ECSN progenitors in binaries that remain bound before explosion at solar metallicity. Points are color-coded based on the envelope classification: hydrogen-rich Type II (red), Type IIb (orange), Type Ib (light blue), and Type Ic (dark blue). The classification thresholds are defined in Section 3.2 based on residual hydrogen and helium envelope masses. The top and right panels show kernel density estimates of the marginal distributions.}
    \label{fig:presn_mass_period}
\end{figure*}

\subsection{Pre-supernova structure, classification, and mass loss}

To assign an approximate envelope-based supernova class to each system, we use the remaining hydrogen and helium envelope masses at collapse. These labels are not light-curve classes derived from radiation transport calculations. Systems retaining a relatively massive hydrogen envelope are grouped as hydrogen-rich Type II events. Type IIb is treated separately as a stripped Type II subtype, following the classical case of SN 1993J, whose early hydrogen features weaken as the spectrum evolves toward a Type Ib-like appearance. Modeling of SN 1993J by \citet{nomoto_type_1993} showed that its Type IIb behavior required a substantially stripped hydrogen envelope. Binary progenitor models for Type IIb and Ib supernovae also show that the remaining hydrogen envelope separates the two classes across a relatively small mass interval \citep{yoon_type_2017,gilkis_how_2022}, so we adopt $0.5\,M_\odot$ as the upper hydrogen envelope mass for Type IIb systems. For the IIb/Ib boundary, \citet{hachinger_how_2012} found from NLTE radiative transfer calculations that $0.01-0.05\,M_\odot$ of hydrogen mixed into the ejecta is sufficient to produce a clear $\rm H\alpha$ absorption feature, so we adopt $0.033\,M_\odot$, close to the middle of this interval, as the dividing value. For the Ib/Ic boundary, \citet{hachinger_how_2012} found that helium masses of $0.06-0.14\,M_\odot$ can produce detectable optical He I lines in their models, motivating our adoption of $0.1\,M_\odot$ as a practical population synthesis threshold. Using these thresholds, we classify systems with $M_{\rm H}>0.5\,M_\odot$ as hydrogen-rich Type II, those with $0.033M_\odot\le M_{\rm H}\le0.5M_\odot$ as Type IIb, those with $M_{\rm H}<0.033\,M_\odot$ and $M_{\rm He}\ge0.1\,M_\odot$ as Type Ib, and the remaining low helium systems as Type Ic.

The joint distribution of pre-supernova orbital period and progenitor mass for primary ECSN progenitors in binaries that avoid merger at solar metallicity is shown in Fig.~\ref{fig:presn_mass_period}(a) as a representative case. Rather than separating into several isolated groups, the population forms a distinct L-shaped band in the pre-supernova mass--period plane. The same classification scheme also includes Type Ic systems, although these are nearly absent in the present population.

The geometry reflects two different pieces of binary physics encoded in the pre-supernova mass and orbital period. Along the nearly vertical branch at $P_{\rm orb,preSN}\simeq2000-3000\,{\rm d}$, the progenitor mass decreases from about $7\,M_\odot$ to about $2.2\,M_\odot$. This branch corresponds to SAGB progenitors that fill their Roche lobes before the ECSN and transfer an increasing fraction of their envelopes to the companion. As the envelope is removed, the pre-supernova mass drops nearly monotonically along this branch. Once the hydrogen envelope is exhausted, the remaining mass is set mainly by the ONe core and residual helium envelope, producing the nearly horizontal Type Ib sequence at $M_{\rm preSN}\simeq2.2\,M_\odot$.

Systems with shorter pre-supernova orbital periods also experienced interaction during the SAGB phase, but their closer orbits caused envelope loss to occur earlier; wider systems can instead retain a more massive envelope until collapse. As a result, systems with $P_{\rm orb,preSN}\simeq50-1000\,{\rm d}$ remain clustered around a similar pre-supernova mass. Only at the shortest periods, $P_{\rm orb,preSN}\lesssim20\,{\rm d}$, does the progenitor mass decrease further to about $1.8\,M_\odot$, indicating additional stripping of the helium-rich layers. A small number of short period systems retain higher pre-supernova masses, which likely reflects reverse mass transfer from the initially less massive companion back onto the ECSN progenitor.

The near absence of Type Ic progenitors follows the same logic. In the surviving binary channel, stripping is efficient enough to remove nearly all of the hydrogen envelope and produce Type Ib systems, but it rarely reduces the helium mass below the adopted $0.1\,M_\odot$ Ib/Ic boundary. Systems with still shorter periods would be expected to experience stronger stripping, but they are also more likely to enter dynamically unstable interaction and merge rather than survive as pre-supernova binaries. Thus, in the present models, Type Ib represents the deepest stripping normally reached by ECSN progenitors that avoid merger, rather than an intermediate stage on a continuous path toward Type Ic.

\begin{figure*}
    \centering
    \includegraphics[width=0.98\textwidth]{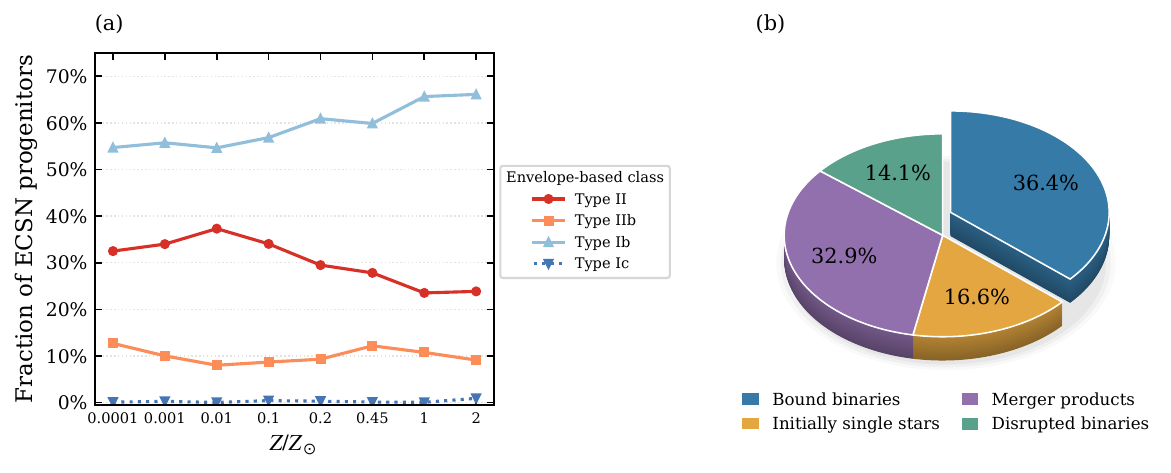}
    \caption{(a) Envelope-based pre-supernova class fractions of ECSN progenitors in binaries that remain bound before explosion, as a function of metallicity. Primary and secondary ECSNe are combined, and the fractions at each metallicity are normalized by the total weighted count of ECSNe in the bound sample. The classes are defined by the residual hydrogen and helium mass thresholds described in Section 3.2. (b) Origins and system states immediately before explosion for all ECSNe at solar metallicity. The four fractions are normalized by the total weighted count of ECSNe at this metallicity and sum to unity. The raised sector denotes bound binaries.}
    \label{fig:spectral_fraction_metallicity}
\end{figure*}

With the adopted classification scheme, the Type IIb population falls at the bend of the L-shaped distribution. This placement makes the marginal distributions especially useful diagnostically. Within the solar metallicity sample of binaries that avoid merger, an ECSN classified as hydrogen-rich Type II is associated with a narrow pre-supernova mass range near $6.7-7.0\,M_\odot$ and predominantly wide orbits, typically $P_{\rm orb,preSN}\gtrsim1000\,{\rm d}$. An ECSN classified as Type Ib, by contrast, implies a much lower pre-supernova mass, concentrated around $1.8-2.2\,M_\odot$, although its orbital period remains less tightly constrained as the Ib population extends over a broad range of periods. Type IIb systems occupy the transition near the bend of the L-shaped distribution, with progenitor masses between the hydrogen-rich Type II and Type Ib peaks and periods overlapping the wide-orbit hydrogen-rich Type II population. Thus, the envelope-based class does not merely encode the stripping depth; in these models it also provides a useful constraint on the pre-supernova mass and, for hydrogen-rich Type II systems in particular, the orbital scale of the binary.

The secondary population in Fig.~\ref{fig:presn_mass_period}(b) does not show the same continuous L-shaped sequence. Hydrogen-rich Type II progenitors are concentrated near $M_{\rm preSN}\simeq6.7-7.0\,M_\odot$ and usually have $P_{\rm orb,preSN}\gtrsim10^3\,{\rm d}$. Type Ib progenitors, by comparison, form a lower-mass group near $1.4-2.4\,M_\odot$ over a broader period range. Across the metallicity grid, the secondary population consists predominantly of Type Ib progenitors that undergo Roche-lobe overflow as donors and Type II progenitors that generally avoid it, with very few intermediate Type IIb outcomes.

Secondary progenitors yield fewer Type IIb outcomes than primary progenitors, partly because the stability of mass transfer differs between the two channels. Primary ECSN progenitors can undergo stable mass transfer to companions of comparable mass, removing most of the hydrogen-rich envelope while leaving a small amount of residual hydrogen. By contrast, when the initially less massive secondary becomes the donor, the initially more massive star has typically already evolved into a compact object. The donor-to-accretor mass ratio can therefore be larger,
particularly for WD or NS accretors, favoring unstable mass transfer and CE evolution \citep{soberman_stability_1997,Nie2025,Nie2026}. Under the CE treatment described in Section~2.1, surviving secondary donors are generally stripped down to their helium cores, leaving little parameter space for partially stripped Type IIb progenitors.

Figure~\ref{fig:spectral_fraction_metallicity}(a) shows the envelope-based class fractions of ECSNe occurring in binaries that remain bound immediately before explosion, combining primary and secondary progenitors. Type Ib systems dominate this bound-binary sample, accounting for approximately $60\%$, while hydrogen-rich Type II and Type IIb systems contribute approximately $30\%$ and $10\%$, respectively; Type Ic systems are nearly absent. Because primary ECSNe dominate the bound-binary sample, the combined class fractions closely follow those of the primary ECSN population.

Toward the metal-rich end of the grid, the Type Ib fraction increases at the expense of the hydrogen-rich Type II fraction. This trend is consistent with the metallicity dependence discussed in Section~3.1. Stronger winds at higher metallicity enhance envelope loss, particularly among the more massive ECSN progenitors, thereby reducing the fraction that retain substantial hydrogen envelopes until collapse. The variation remains moderate because the final envelope-based class is governed primarily by the timing and depth of binary stripping. Metallicity-dependent winds therefore modulate the balance between hydrogen-rich Type II and Type Ib outcomes without erasing the structure produced by binary interaction in Fig.~\ref{fig:presn_mass_period}(a).

Figure~\ref{fig:spectral_fraction_metallicity}(b) shows that, at solar metallicity, $36.4\%$ of ECSNe occur in binaries that remain bound immediately before explosion, while $16.6\%$ arise from initially single stars, $32.9\%$ from merger products, and $14.1\%$ from stars whose binaries were disrupted by an earlier supernova. The bound-binary contribution ranges from $33.2\%$ to $37.9\%$ across the metallicity grid. Progenitors in the disrupted-binary category were generally born as the initially less massive components; however, they are single stars at the time of their ECSNe and are therefore classified as disrupted-binary events rather than as secondary ECSNe in bound systems. Most ECSNe consequently occur without a bound companion, even though many originate from binary systems. The three unbound categories are dominated by hydrogen-rich Type II events, reflecting their comparatively limited envelope stripping before collapse.

Combining the origin fractions with the envelope-based class fractions within each group, hydrogen-rich Type II events account for approximately $72\%-77\%$ of the full ECSN population, Type IIb events for $3\%-4\%$, Type Ib events for $19\%-24\%$, and Type Ic events for less than $0.4\%$. Including the numerous hydrogen-rich events without bound companions therefore changes the dominant class from Type Ib in the bound-binary sample to hydrogen-rich Type II in the full ECSN population.

\begin{figure*}
    \centering
    \includegraphics[width=0.9\textwidth]{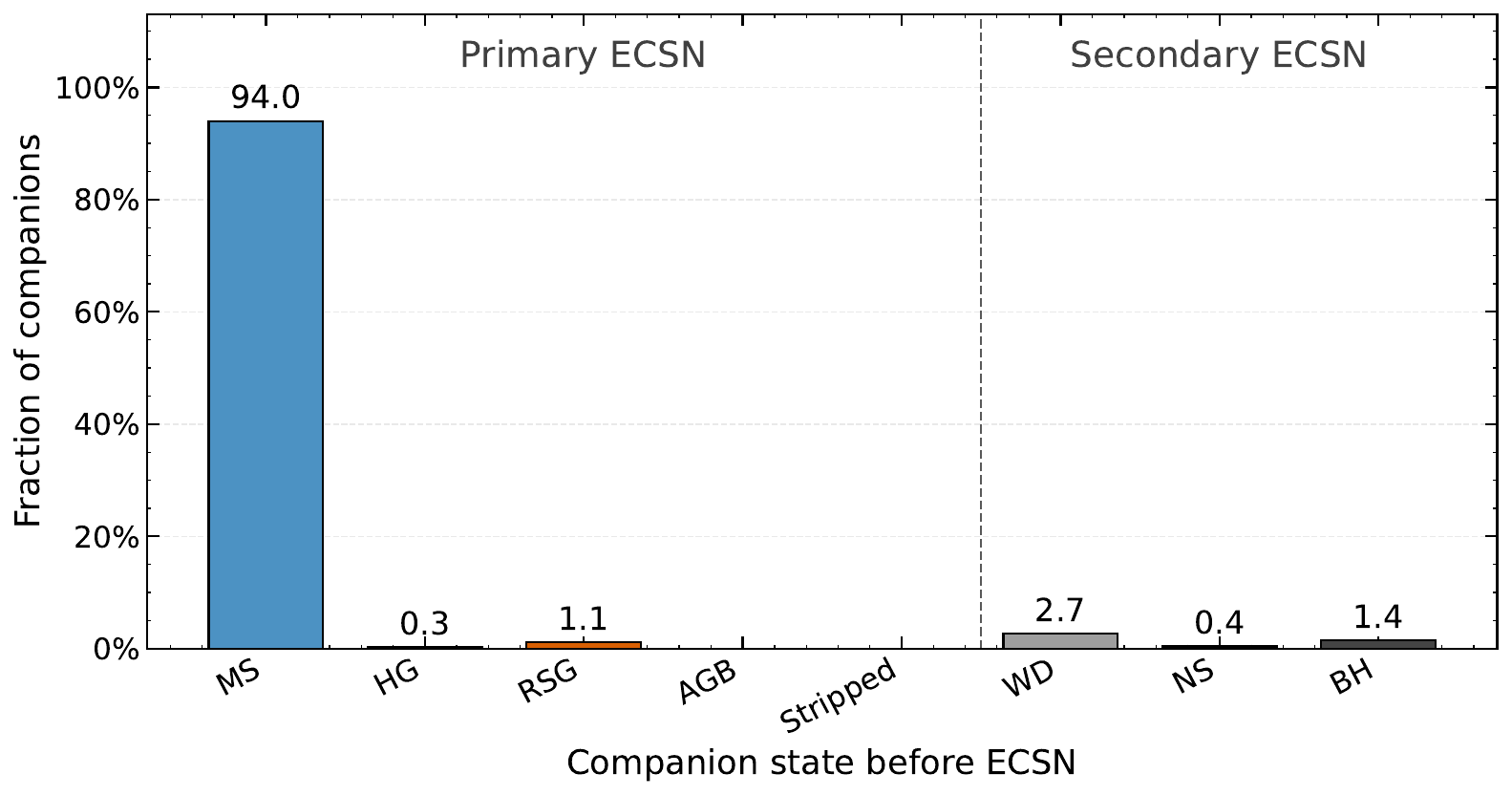}
    \caption{Evolutionary states of companions immediately before ECSNe in bound binaries at solar metallicity. The dashed line separates primary ECSNe (left) from secondary ECSNe (right). All bars are normalized by the total weighted count of primary and secondary ECSNe in the bound-binary sample, and the percentages give the corresponding fractions. MS, HG, RSG, AGB, WD, NS, and BH denote main sequence, Hertzsprung gap, red supergiant, asymptotic giant branch, white dwarf, neutron star, and black hole, respectively. Stripped denotes a stripped-star companion.}
    \label{fig:companion_state}
\end{figure*}

HD~45166 is particularly relevant in this context because it hosts a magnetic helium star with a mass of approximately $2\,M_\odot$, which has been proposed as a possible future ECSN progenitor. \citet{shenar_massive_2023} suggested that this helium star formed from an initially more massive star that was first stripped and subsequently received material through reverse mass transfer from its initially less massive companion after the latter had evolved into a giant. This interaction led to CE evolution and ultimately to a merger. Our solar-metallicity population includes systems with broadly analogous pre-merger configurations. At the onset of the CE phase, the initially more massive star is undergoing core helium burning and is nearly fully stripped, while the companion is a giant. However, the approximate CE-merger prescription adopted in POSYDON does not resolve the detailed coalescence of a stripped star with a giant. In particular, it retains most of the hydrogen-rich envelope, whereas the scenario proposed by \citet{shenar_massive_2023} ejects most of this material during the merger. Consequently, our models do not produce a merger remnant resembling the $\sim2\,M_\odot$ helium star in HD~45166. Possible formation pathways for HD~45166 and related systems have also been explored by \citet{Picco2026} and \citet{Wang2026b}.

\subsection{Pre-supernova companions and post-supernova systems}

We first examine the evolutionary state of the companion immediately before the ECSN, restricting the analysis to binaries that remain bound at the time of explosion. Fig.~\ref{fig:companion_state} shows the distribution for the solar-metallicity sample. Primary ECSNe account for $95.46\%$ of the combined bound sample; systems with a main-sequence companion alone account for $94.00\%$. Hertzsprung-gap and red-supergiant companions are rare, contributing only about $0.34\%$ and $1.12\%$, respectively, while AGB and stripped-star companions are nearly absent. The remaining $4.54\%$ are secondary ECSNe, all of which have compact companions in this sample: WDs ($2.71\%$), neutron stars ($0.40\%$), and black holes ($1.43\%$).

The predominance of main-sequence companions follows from the ordering of the nuclear timescales. When the initially more massive star undergoes a primary ECSN, its initially less massive companion is usually still undergoing core hydrogen burning. Producing an evolved companion at this stage requires an initial mass close to that of the primary, so that the two stars evolve on comparable nuclear timescales; the binary must also avoid merging before the primary collapses. These requirements leave only a small fraction of systems with Hertzsprung-gap or red-supergiant companions at the time of the primary ECSN. By contrast, the companions of bound secondary-ECSN progenitors in this sample have already evolved into compact remnants.

The relative abundances of WD, neutron-star, and black-hole companions also reflect whether the binary survives the earlier evolution of the primary. WD formation imparts no natal kick in our models, whereas the stronger kicks assigned to neutron stars formed in Fe CCSNe can disrupt the system before the secondary reaches its ECSN. In the adopted prescription, black-hole kick velocities are scaled inversely with remnant mass, thereby increasing the survival probability of binaries containing black holes relative to those containing neutron stars formed in Fe CCSNe.

Table~\ref{tab:companion_states} extends this classification across the full metallicity grid. Primary ECSNe with main-sequence companions account for $88.1\%-95.8\%$ of the combined bound-binary sample, whereas secondary ECSNe contribute $2.3\%-8.2\%$; WDs are their most common companions. The companion-state fractions of primary ECSNe vary only weakly with metallicity: main-sequence companions dominate at every $Z$, while Hertzsprung-gap, red-supergiant, AGB, and stripped-star companions remain minor channels. The broad companion demographics are therefore governed mainly by the relative evolutionary timescales of the two stars, with metallicity-dependent winds playing a secondary role. Thus, the most likely stellar counterpart to a primary ECSN is a main-sequence companion, although previous mass transfer may have altered its mass, surface composition, and rotation relative to those of an isolated star of the same age.

\begin{table*}
    \centering
    \caption{Companion-state fractions immediately before ECSNe in bound binaries as a function of metallicity. Columns are grouped by the ECSN progenitor: primary or secondary. At each metallicity, all entries are normalized to the total weighted count of primary and secondary ECSNe in the bound-binary sample.}
    \label{tab:companion_states}
\begin{tabular}{lrrrrrrrr}
\toprule
 & \multicolumn{5}{c}{Primary ECSN} & \multicolumn{3}{c}{Secondary ECSN} \\
\cmidrule(lr){2-6}\cmidrule(lr){7-9}
$Z/Z_\odot$ & MS & HG & RSG & AGB & Stripped & WD & NS & BH \\
\midrule
0.0001 & 92.243\% & 2.751\% & 1.167\% & 0.042\% & 0.000\% & 2.626\% & 0.234\% & 0.936\% \\
0.001 & 91.269\% & 1.810\% & 2.610\% & 0.042\% & 0.000\% & 2.361\% & 0.963\% & 0.945\% \\
0.01 & 88.131\% & 1.022\% & 2.640\% & 0.043\% & 0.000\% & 6.926\% & 0.734\% & 0.505\% \\
0.1 & 91.847\% & 0.846\% & 2.490\% & 0.047\% & 0.000\% & 3.461\% & 0.810\% & 0.498\% \\
0.2 & 95.786\% & 0.591\% & 1.330\% & 0.000\% & 0.000\% & 1.365\% & 0.375\% & 0.553\% \\
0.45 & 94.018\% & 0.348\% & 1.936\% & 0.000\% & 0.000\% & 2.192\% & 0.329\% & 1.177\% \\
1 & 93.995\% & 0.341\% & 1.120\% & 0.000\% & 0.000\% & 2.715\% & 0.402\% & 1.428\% \\
2 & 93.199\% & 0.267\% & 2.187\% & 0.360\% & 0.107\% & 2.150\% & 0.567\% & 1.165\% \\
\bottomrule
\end{tabular}

\end{table*}

\begin{figure}
    \centering
    \includegraphics[width=0.92\columnwidth]{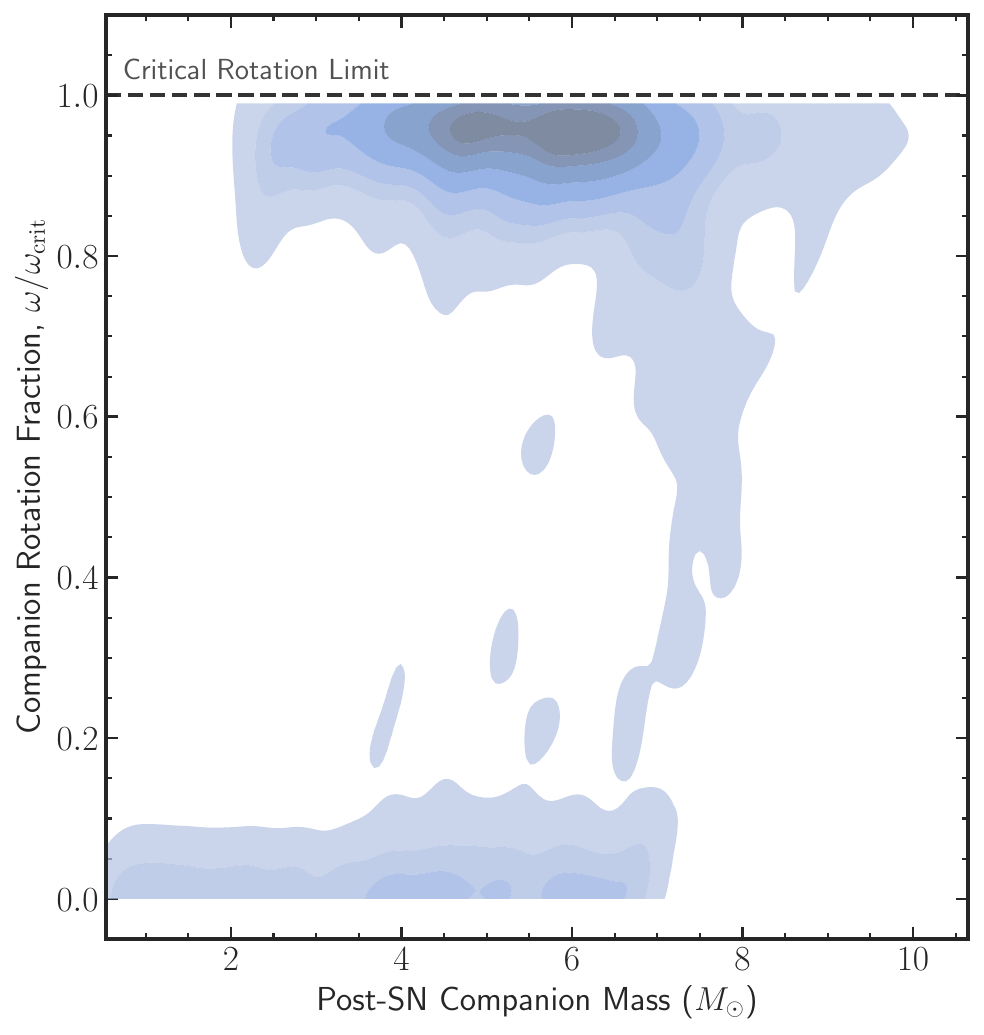}
    \caption{Post-supernova companion mass and surface rotation after a primary ECSN at solar metallicity. The vertical axis shows the companion's surface rotation rate normalized by the critical value, $\omega/\omega_{\rm crit}$, and the dashed horizontal line marks critical rotation ($\omega/\omega_{\rm crit}=1$). Color contours indicate the kernel density distribution of surviving companions in the mass--rotation plane.}
    \label{fig:companion_spin}
\end{figure}

Mass accretion during close binary evolution can spin up companions efficiently; in the classical limit, only a modest amount of accreted matter is needed to drive a star toward critical rotation \citep{packet_spin-up_1981}. As shown in Fig.~\ref{fig:companion_spin}, the surviving companions after the primary ECSN at solar metallicity form a bimodal distribution in $\omega/\omega_{\rm crit}$, where $\omega$ is the companion's angular velocity and $\omega_{\rm crit}$ is its critical value. One component remains slowly rotating, with $\omega/\omega_{\rm crit}\lesssim0.1$ across a broad mass range. The second component forms a dense near-critical branch at $\omega/\omega_{\rm crit}\simeq0.9-1.0$, most prominently for companion masses of $3-7\,M_\odot$, with the highest density around $5-6\,M_\odot$. This mass range for Be stars is dependent on the treatment of mass-transfer efficiency during progenitor binary evolution \citep[e.g.,][]{Shao2014,Rocha2024,Xu2025,Schurmann2025,Wang2026c,LopezOller2026}. The POSYDON grids used here limit further accretion as the receiving star approaches critical rotation. \citet{xing_disk_regulated_2026} proposed an alternative model in which disks regulate angular-momentum transfer, allowing mass growth to continue near critical rotation and producing more massive Be stars.

Fig.~\ref{fig:companion_spin} further shows that rotation divides the mostly main-sequence companion population into two interaction histories. Wider systems, or systems in which the companion accreted inefficiently, remain on the slow rotation branch, while systems that underwent active RLOF can spin the companion up to near-critical rotation. This rotational information is not visible in the companion state census alone, where most stellar companions in Fig.~\ref{fig:companion_state} are still in core hydrogen burning at the time of primary ECSN.

The near-critical branch is also relevant for the possible connection between ECSNe and rapidly rotating companions in Be/X-ray binaries \citep[e.g.,][]{Shao2014}. Rotation alone, however, is not sufficient to identify such descendants, as the post-supernova orbital period and eccentricity are also shaped by mass loss and the natal kick.


\begin{figure}
    \centering
    \includegraphics[width=0.98\columnwidth]{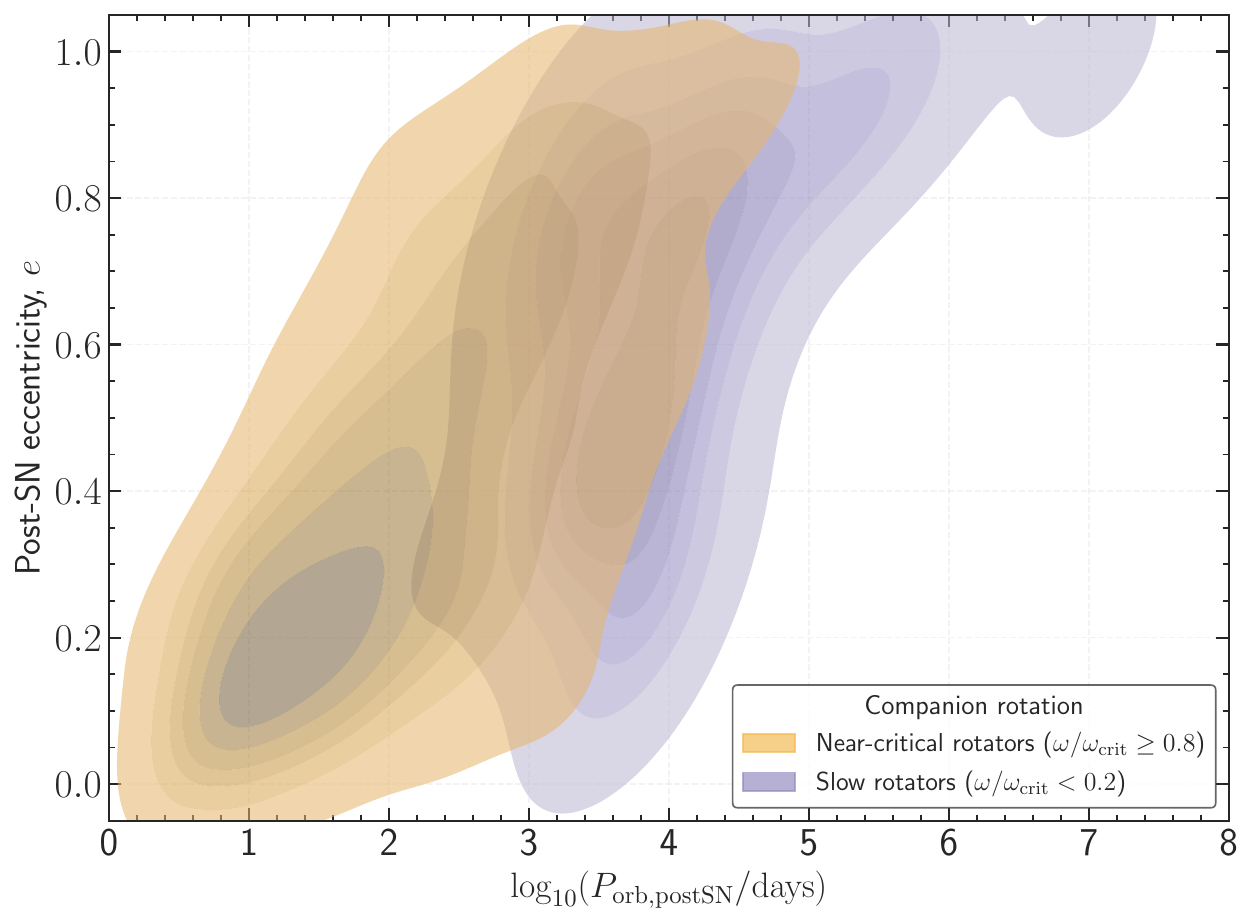}
    \caption{Post-supernova orbital period and eccentricity of binaries that survive a primary ECSN at solar metallicity, distinguished by the rotation rate of the companion star. Orange contours denote near-critical rotators with $\omega/\omega_{\rm crit}\ge0.8$, and purple contours represent slowly rotating companions with $\omega/\omega_{\rm crit}<0.2$.}
    \label{fig:postsn_orbits_rotation}
\end{figure}

These distinct spin branches correlate with the post-supernova orbital configurations shown in Fig.~\ref{fig:postsn_orbits_rotation}. The two populations overlap, but the near-critical branch is shifted as a whole toward shorter periods and lower eccentricities relative to the slow rotation branch. Near-critical companions have their strongest concentration at short periods, $P_{\rm orb,postSN}\sim10-100\,{\rm d}$, and small eccentricities, $e\simeq0.1-0.3$. Slowly rotating companions peak at much wider periods, roughly $P_{\rm orb,postSN}\sim10^3-10^5\,{\rm d}$, with typical eccentricities of $e\simeq0.4-0.9$, and their low density tail extends to periods above $10^6\,{\rm d}$ and eccentricities close to $e=1$.

The displacement between the two spin branches reflects the pre-supernova orbital scale. Near-critical rotators usually come from tighter systems, where stronger interaction spins up the companion before the primary ECSN. Their post-supernova orbits inherit this compact scale, and the larger orbital velocities make a given ECSN kick less effective at increasing the eccentricity. Slow rotators more often come from weakly interacting or wider binaries, giving them a broader period range and higher typical eccentricities. The CE survivors form a small exception to this trend. Their pre-supernova periods are relatively short, with a typical scale of order $10\,{\rm d}$, but they account for only about $1\%$ of the non-merged binaries across the full metallicity grid. This small fraction keeps them from setting the density contours in Figs.~\ref{fig:companion_spin} and~\ref{fig:postsn_orbits_rotation}.

After the interaction history has set the pre-supernova orbit, the final eccentricity is further shaped by the mass lost during neutron star formation and by the natal kick. As a useful analytic reference, symmetric instantaneous mass loss from an initially circular binary produces a Blaauw recoil \citep{blaauw_origin_1961}. In this limiting case, the system remains bound when the mass lost ($\Delta M$) is less than half of the total pre-supernova mass  ($M_1+M_2$), and the post-supernova eccentricity satisfies $e=\Delta M/(M_1+M_2-\Delta M)$ when the kick is negligible. The POSYDON models used here also include natal kicks, which broaden this simple relation \citep{hills_effects_1983}. The effect of a given kick is set by its size relative to the pre-supernova orbital velocity. Compact systems, which usually have larger orbital velocities, are less perturbed by the same ECSN kick than wider systems with slower orbital motion. This helps explain why the near-critical branch, already associated with shorter pre-supernova and post-supernova periods, is weighted toward lower eccentricities. Given the relatively low ejecta masses and weak kicks expected for ECSNe, such compact systems can remain bound with smaller eccentricities more easily than comparable systems undergoing Fe CCSNe \citep{pfahl_new_2002,van_den_heuvel_high_2010,tauris_formation_2017}.

Combining the near-critical companion rotation in Fig.~\ref{fig:companion_spin} with the compact, low-eccentricity region in Fig.~\ref{fig:postsn_orbits_rotation} gives a natural route to some Be/X-ray binaries. Rapid rotation is a defining ingredient of Be star phenomenology, while low kicks help preserve bound systems with small eccentricities. Observed Be/X-ray binaries have also been interpreted as containing a short-period, low-eccentricity subpopulation preferentially associated with ECSNe \citep[][]{knigge_two_2011}. The near-critical branch in Figs.~\ref{fig:companion_spin} and~\ref{fig:postsn_orbits_rotation} provides a plausible ECSN pathway to this class of systems, although the formation of an observable X-ray binary also requires subsequent wind accretion or RLOF \citep{Cheng2014}.

\subsection{Cosmic ECSN rates across cosmic time}

To obtain the cosmic ECSN rate, we combine the ECSN efficiency as a function of metallicity and delay time with three prescriptions for the cosmic star formation history and the metallicity distribution of newly formed stars. For the ECSN progenitors considered here, the delay from ZAMS to collapse is longer than for the most massive Fe CCSN progenitors but still short on cosmological timescales, making the intrinsic ECSN rate closely track the cosmic star formation history after weighting by metallicity.

We calculate the intrinsic cosmic rate using the POSYDON Rates module, which combines the simulated formation efficiency with a chosen cosmic star formation and metallicity distribution model. In the metallicity-resolved star formation framework,
the intrinsic ECSN rate density can be written as
\begin{equation}
    \hspace*{\fill}
    \mathcal{R}_{\rm ECSN}(z)=\int dZ\int dt_{\rm d}\,\eta_{\rm ECSN}(Z,t_{\rm d})\,\dot{\rho}_\star[Z,z_{\rm f}(z,t_{\rm d})]
    \hspace*{\fill}
    \label{eq:cosmic_ecsn_rate}
\end{equation}
where \(\dot{\rho}_\star(Z,z_{\rm f})=\dot{\rho}_\star(z_{\rm f})p(Z|z_{\rm f})\) denotes the star formation rate density resolved by metallicity at the formation redshift, and \(\eta_{\rm ECSN}(Z,t_{\rm d})\) is the total ECSN efficiency per unit formed stellar mass for each metallicity and delay time; it includes events from either component of a binary, initially single stars, and merger products. Here \(t_{\rm d}\) is the time from ZAMS to collapse. We compare three cosmic prescriptions for the metallicity-dependent star formation history: the empirical fit from \citet{madau_radiation_2017}, the IllustrisTNG/TNG100-1 cosmological simulation \citep{nelson_illustristng_2019}, and the parametric prescription of \citet{neijssel_effect_2019}. To quantify the ECSN contribution relative to the total core-collapse supernova population, we compute \(f_{\rm ECSN}=\mathcal{R}_{\rm ECSN}/\mathcal{R}_{\rm SN}\), where \(\mathcal{R}_{\rm SN}\) is the total simulated rate of ECSNe and Fe CCSNe.

\begin{figure}
    \centering
    \includegraphics[width=0.96\columnwidth]{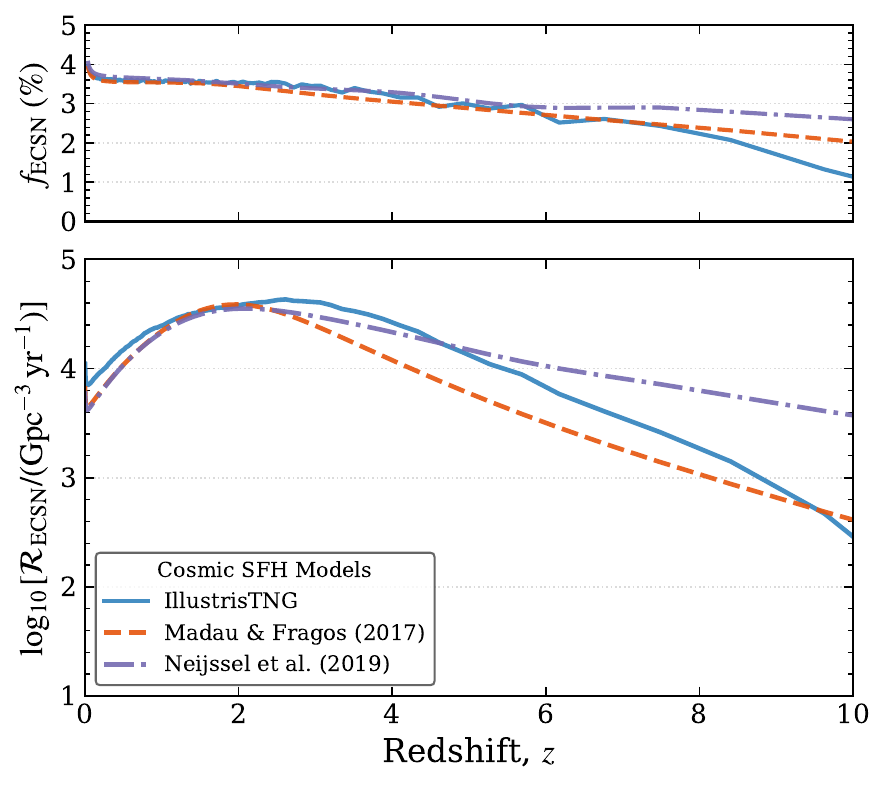}
    \caption{Intrinsic comoving rate density of all ECSNe (bottom) and their fraction relative to all simulated supernovae (top) as a function of redshift. The ECSN sample includes events from both components of binaries, initially single stars, and merger products. Curves show three cosmic models: IllustrisTNG/TNG100-1 \citep{nelson_illustristng_2019}, \citet{madau_radiation_2017}, and \citet{neijssel_effect_2019}. The ECSN fraction is defined as \(f_{\rm ECSN}=\mathcal{R}_{\rm ECSN}/\mathcal{R}_{\rm SN}\)}.
    \label{fig:cosmic_rates}
\end{figure}

As shown in Fig.~\ref{fig:cosmic_rates}, all three models place the rate maximum at similar redshift. At \(z\simeq0\), the intrinsic ECSN rate is
\(\sim(4-7)\times10^3\,{\rm Gpc}^{-3}\,{\rm yr}^{-1}\).
It rises to \(\sim(3.5-4.3)\times10^4\,{\rm Gpc}^{-3}\,{\rm yr}^{-1}\)
at \(z\simeq2\)--3 and then declines at higher redshift. The peak coincides with the broad maximum of the cosmic star formation rate density. The corresponding ECSN fraction is a few per cent over the plotted redshift range, decreasing from \(\sim4\%\) at low redshift to  \(\sim1\%-3\%\) by \(z\simeq10\), depending on the adopted cosmic model.
At \(z\simeq0\), both the predicted rate and the ECSN fraction are broadly consistent with recent estimates based on an observed sample of ECSN candidates \citep{Sato2026}.

The three cosmic prescriptions differ most near the star formation peak and in the high redshift tail. IllustrisTNG gives the largest ECSN rate around $z\sim2-3$, while the \citet{madau_radiation_2017} model falls more steeply after the peak. The \citet{neijssel_effect_2019} prescription retains more weight at high redshift and gives the largest rate for $z\gtrsim5$. The fractional contribution \(f_{\rm ECSN}\) changes less than \(\mathcal{R}_{\rm ECSN}\) at low and intermediate redshift, indicating that much of the evolution is shared by ECSNe and the wider supernova population. At higher redshift, the result becomes more sensitive to the assumed metallicity distribution, as the ECSN channel samples a narrow progenitor window whose efficiency changes across the metallicity grid.

\FloatBarrier

\section{Conclusions}

We have used POSYDON to study binary ECSN progenitors across metallicity, focusing on how the narrow internal structure criterion for ECSNe maps into initial binary parameter space, pre-supernova observables, surviving companions, and cosmic rates. The main results are summarized as follows.

(i) Owing to efficient convective overshooting that increases with mass in the POSYDON/MESA grids, the ECSN-forming primary mass window is shifted to lower masses than the classical single-star expectation. Stars around $6.5-7\,M_\odot$ can develop CO cores in the \citet{tauris_ultra-stripped_2015} ECSN range, setting the lower boundary of the population. This boundary varies only weakly with metallicity over most of the grid, while the upper boundary is more sensitive to $Z$ and moves to higher initial masses in metal-rich environments, where stronger winds alter the envelope and core growth history.

(ii) When this mass interval is mapped back onto the initial binary parameter space, ECSN formation is controlled by the timing and depth of envelope removal. Binaries that avoid merger occupy a continuous region in the $M_{1,\rm i}$--$P_{\rm orb,i}$ plane for higher mass ratios, whereas low mass-ratio systems show a larger merger contribution. The merger component should be interpreted qualitatively, given the approximate treatment of merger products in POSYDON, but it indicates that unstable interaction can contribute to the ECSN-forming population.

(iii) In the pre-supernova mass--period plane, primary ECSN progenitors in binaries that avoid merger form an L-shaped distribution. The envelope-based classification of ECSNe places Type II systems on the hydrogen-rich branch, Type Ib systems on the stripped low-mass branch, and Type IIb systems near the bend of the distribution. This location gives the IIb class diagnostic value, as it corresponds to a relatively restricted region of pre-supernova mass and orbital period. Type Ic systems are nearly absent, indicating that stripping in non-merged binaries usually removes hydrogen without reducing the helium mass below the adopted Ib/Ic threshold. Bound secondary ECSN progenitors show a sharper division between a hydrogen-rich Type II group at high mass and wide orbital period and a stripped Type Ib group at lower mass, with few systems occupying intermediate envelope classes. Events without a bound companion are predominantly hydrogen rich, making Type II the most common class in the full ECSN population, with a fraction of $\sim72\%-77\%$.

(iv) Binaries bound before explosion account for $33\%-38\%$ of all ECSNe across the metallicity grid. For the companion population, most systems that retain a stellar companion at primary ECSN contain a main-sequence star, while evolved companions remain rare across the metallicity grid. Secondary ECSNe contribute $2.3\%-8.2\%$ of the combined bound sample and have compact companions. This contrast follows from the evolutionary ordering: at primary ECSN, the initially less massive star usually remains in core hydrogen burning, whereas by the time the secondary reaches its ECSN, its companion has already become a compact remnant. Companion rotation provides a stronger imprint of previous interaction than the companion evolutionary state alone. The surviving companions after the primary ECSN at solar metallicity show a bimodal spin distribution, with a slow rotation branch and a near-critical branch associated with prior mass transfer. When mapped onto the post-supernova orbital plane, the near-critical branch is shifted toward shorter periods and lower eccentricities relative to the slow rotation branch, providing a possible link between ECSNe, spun-up companions, and some low-eccentricity Be/X-ray binaries.

(v) After convolution with cosmic star formation and metallicity evolution models, the intrinsic ECSN rate rises from the local Universe to a maximum around $z\sim2-3$ and then decreases toward earlier cosmic times. ECSNe contribute $1\%-4\%$ of all core-collapse supernova events across most of the plotted redshift range. At $z\gtrsim5$, the spread among the three cosmic prescriptions is driven by how each model assigns high-redshift star formation to low-metallicity bins. The \citet{neijssel_effect_2019} model retains a larger low-metallicity contribution through its Langer--Norman metallicity prescription, whereas the \citet{madau_radiation_2017} prescription combines an analytic cosmic star formation history with an empirical mean metallicity relation, giving less weight to the lowest metallicity bins at high redshift and producing a faster decline in the ECSN rate. IllustrisTNG instead follows the tabulated star formation and chemical enrichment history of the cosmological simulation.

Taken together, these results show that binary ECSNe are shaped by a combination of internal mixing, envelope stripping, companion accretion, and supernova dynamics. Future work improving the treatment of mergers and comparing with observed ECSN candidates and X-ray binary populations will help test the pathways identified here.

\section*{Acknowledgements}

We thank the referee for constructive suggestions that helped improve this paper. This work was supported by the National Key Research and Development Program of China (Grant Nos. 2023YFA1607902, 2021YFA0718500).

\section*{Data Availability}
The population synthesis outputs generated in this study will be shared on reasonable request to the corresponding author.

\bibliographystyle{mnras}
\bibliography{references}

\label{lastpage}
\end{document}